\documentclass[11pt]{article}
\usepackage{amsmath,amsthm,latexsym,amssymb,amsfonts,epsfig}

\newtheorem{lemm}{Lemma}[section]
\newtheorem{prop}[lemm]{Proposition}
\newtheorem{coro}[lemm]{Corollary}
\newtheorem{defi}[lemm]{Definition}

\newtheorem{theo}[lemm]{Theorem}

\newcommand{\hilb}{\mathcal{H}}

\newcommand{\one}{\text{\bf 1}}

\newcommand{\gcon}{\overline{\mathcal{A}}}
\newcommand{\con}{\mathcal{A}}

\newcommand{\rep}[1]{\pi\left(#1\right)}

\newcommand{\scpr}[2]{\left\langle#1\,,\, #2 \right\rangle} 

\newcommand{\norm}[1]{\left\lVert #1 \right\rVert}     
\newcommand{\betr}[1]{\left\lvert #1 \right\rvert}    
\newcommand{\mal}{\mu_{\text{AL}}}
\newcommand{\mh}{\mu_{\text{H}}}

\newcommand{\avec}[1]{\underline{#1}}
\newcommand{\R}{\mathbb{R}}              
\newcommand{\Z}{\mathbb{Z}}              
\newcommand{\C}{\mathbb{C}}

\newcommand{\qqquad}{\qquad\qquad}

\DeclareMathOperator{\porder}{\mathcal{P}}

\DeclareMathOperator{\cyl}{Cyl}
  
\DeclareMathOperator{\add}{{\bf add}}  
\DeclareMathOperator{\sub}{{\bf sub}}

\DeclareMathOperator{\sutwo}{SU(2)}
\DeclareMathOperator{\uone}{U(1)}

\begin{document}
\title{When Do Measures on the Space of Connections Support 
the Triad Operators of Loop Quantum Gravity?}
\author{Hanno Sahlmann\thanks{sahlmann@apctp.org. This work was done while 
at the MPI f\"ur Gravitationsphysik, Albert-Einstein-Institut, Potsdam, Germany}\\
{\small Asia Pacific Center for Theoretical Physics, Pohang (Korea)}}
\date{{\small PACS No. 04.60, Preprint AEI-2002-057}}

\maketitle

\begin{abstract}
In this work we investigate the question, under what conditions  
Hilbert spaces that are induced by measures on the space of
generalized connections carry a representation of certain non-Abelian 
analogues of the electric flux.
We give the problem a precise mathematical formulation and start its
investigation. 
For the technically simple case of U(1) as gauge group, we establish 
a number of ``no-go theorems'' asserting that for certain classes of
measures, the flux operators can not be represented on the
corresponding Hilbert spaces.  

The flux-observables we consider play an important role in loop
quantum gravity since they can be defined without recurse to a
background geometry, and   
they might also be of interest in the general context of
quantization of non-Abelian gauge theories.  
\end{abstract}
%---------------------------------------------------------------------------
\section{Introduction}
%---------------------------------------------------------------------------
Loop quantum gravity (LQG for short) is a promising approach to the
problem of finding a quantum theory of gravity, and has led to many
interesting insights (see \cite{Thiemann:2001yy} for an extensive and 
\cite{Rovelli:1998yv} for a
shorter non-technical review). It is based on the formulation of gravity as a
constrained canonical system in terms of the Ashtekar variables
\cite{Ashtekar:1986yd}, a canonical pair of an SU(2)-connection $A$ (in
its real formulation) and a triad field $E$ with a nontrivial 
density weight. Both of these take values on a spacial slice $\Sigma$
of the spacetime. 
A decisive advantage of these new variables is that both connection and
triad allow for a metric-independent way of integrating them to form 
more regular functionals on the classical phase space and hence make a 
quantization feasible:\newline
Being a one-form, $A$ can be integrated
naturally (that is, without recurse to background structure) along
differentiable curves $e$ in $\Sigma$, to form \textit{holonomies}
\begin{equation}
\label{eq1.2}
h_e[A]=\porder\exp\left[i\int_e A \right] \qquad \in \sutwo. 
\end{equation}
The density weight of $E$ on the other hand is such that, using an
additional real co-vector field $f^i$ it can be naturally integrated
over oriented surfaces $S$ to form a quantity
\begin{equation}
\label{eq1.1}
E_{S,f}[E]= \int_S f^i(*E)_i
\end{equation}
analogous to the electric flux through $S$. Since the variables
$h_e[A]$ and $E_{S,f}[E]$ do not rely on any background geometry for
their definition, they are very natural in the context of
diffeomorphism invariant theories, and many important results of LQG 
such as the quantization of area and
volume are related to the choice of these variables as basic
observables. With this choice, however, LQG is in
sharp contrast to the usual formulation of (quantum) gauge theories in 
which it is assumed that only when integrated over three or even four
dimensional regions in the spacetime, the quantum fields make sense as 
operators on some Hilbert space. 

All of this makes it worthwhile, to study the representation theory of
the observables \eqref{eq1.2},\eqref{eq1.1} in somewhat general terms. 
Indeed, the representations of the algebra of holonomies
\eqref{eq1.2} is well studied and powerful mathematical tools have
been developed \cite{Ashtekar:1992kc,Ashtekar:1995wa,Ashtekar:1995mh,
Baez:1994id}. It turns out that cyclic representations are
in one-to-one correspondence with measures on the space $\gcon$ of
(generalized) connections. We will briefly review some results of 
these works in Section  \ref{se2}.   
About the representations of the flux variables on the other hand, not 
so much is known. Therefore,
in \cite{1} we considered the representation theory of the holonomies
(or more precisely, a straightforward generalization, the cylindrical
functions) \textit{together} with the momentum variables $E_{S,f}$ in rather
general terms. In the present paper, we continue this work by
focusing on a specific aspect: 
Given a cyclic representation of the cylindrical functions, is it 
possible to also represent the momentum variables on the same Hilbert space? 
It is well known that this is possible for a specific measure on the
space of generalized $\sutwo$ connections, the Ashtekar-Lewandowski
measure $\mal$. It is distinguished by its simple and elegant
definition and by its invariance under diffeomorphisms of the spacial
slice $\Sigma$. The representation it induces is therefore
considered as the fundamental representation for LQG. 

We will see below, that at least in the somewhat simpler case
when the gauge group is $\uone$ instead of $\sutwo$, the
Ashtekar-Lewandowski representation is not only distinguished, it is
unique: $\mal$ induces the only diffeomorphism invariant cyclic
representation of the algebra 
of cylindrical functions which also carries a representation of the
flux observables $E_{S,f}$. A more general result of this kind was recently established in \cite{Lewandowski:2005jk}. 

Recently it became evident that to investigate the semiclassical
regime of LQG it may be useful to also study representations that are 
not diffeomorphism invariant but encode information about a given
classical background geometry. Interesting representations of this
type for the 
algebra of cylindrical functions were discovered in \cite{V2} for the
case of $\uone$ as gauge group and a suitable generalization for the 
$\sutwo$ case was proposed in \cite{Ashtekar:2001xp}. However, 
these representations
do not extend to representations of the cylindrical functions
\textit{and} the flux observables $E_{S,f}$. 
The original motivation of the present work was to remedy this and construct
representations of both, cylindrical functions and fluxes which are
different from the AL-representation.  
This turned out to be very difficult, however. Quite contrary to our
original goal, the results of the present work show that the
constraints put on by requiring the observables $E_{S,f}$ to be
represented are quite tight, and that consequently it is hard to come 
up with such a
representation that is different from the Ashtekar-Lewandowski
representation. 

Since the purpose of the present work is to explore the territory, our
results are mostly concerned with the case of $\uone$ as gauge group. 
This case is technically much less involved than that of a general compact
gauge group because the representation theory of $\uone$ is so
simple. We expect, however, that generalizations of the results 
to other compact gauge groups are possible. 

Our main results for the $\uone$ case are the following: 
\begin{itemize}
\item There is no diffeomorphism invariant measure allowing for a 
  representation of the flux observables other then $\mal$.  
\item The $r$-Fock measures, as well as any other measure obtained by
  ``importing'' a regular Borel measure on the space of Schwartz
  distributions to $\gcon_{\uone}$ with Varadarajan's method do not
  support a representation of the flux observables. 
\item Any measure which is ``factorizing'' in a certain technical
  sense will, if it supports a representation of the flux observables, 
  be very close to $\mal$. Moreover, the only such measure which
  additionally is Euclidean invariant, is $\mal$.  
\end{itemize}
After work on this paper was finished, new results were obtained that
generalize the first point in our list considerably, using ideas
presented here and in \cite{1}. The interested
reader may want to take a look at
\cite{Okolow:2003pk},\cite{Sahlmann:2003in}. 

Let us finish the introduction with a description of the rest of the 
present work:

In the next section, we prepare the ground by briefly reviewing the
projective techniques that are used to define measures on
$\gcon$. Also, we give a description of these measures which will be
used in establishing our results. 

In Section \ref{se3} we state and explain a necessary
and sufficient condition for a measure $\mu$ on $\gcon$ to allow for a 
representation of the flux observables.

Section \ref{se9} serves to investigate the condition found in Section 
\ref{se3} in detail in the case the gauge group is $\uone$. 
First we introduce the notation necessary for this special case and also
specialize the condition. We then proceed to establishing our
main results. 

With section \ref{se4} we close this work by discussing interpretation 
and possible consequences of our results and point out problems left
open. 
%---------------------------------------------------------------------------
\section{Measures on the space of generalized connections}
%---------------------------------------------------------------------------
\label{se2}
We will start by briefly reviewing the projective techniques
\cite{Ashtekar:1995wa,Ashtekar:1995mh} 
which can be used to construct measures on the space of 
connections. Using these methods we then introduce a rather explicit
representation for such measures which we use in the sequel.

Let $\Sigma$ be a three dimensional, connected, analytic manifold. 
\begin{defi}
By an (oriented) \textit{edge} $e$ in $\Sigma$ we shall mean an
equivalence class of analytic maps $[0,1]\longrightarrow \Sigma$,
where two such maps are considered equivalent if they differ by an orientation
preserving analytic reparametrization.
 
A \textit{graph} in $\Sigma$ is defined to be a union of edges such that 
two distinct ones intersect at most in their endpoints.
The endpoints of the edges contained in the graph will be referred to
as its \textit{vertices}, and we will denote the set of edges of a
graph $\gamma$ by $E(\gamma)$. 
\end{defi}
Analyticity of the edges is required to exclude certain pathological 
intersection structures of the edges with surfaces which would render
the Poisson brackets which will be introduced below ill-defined.

The set of graphs can be endowed with a partial order $\geq$ by
stating that $\gamma'\geq\gamma$ whenever $\gamma$ is contained in
$\gamma'$ in the sense that each edge of $\gamma$ can be obtained as
composition of edges of $\gamma'$ and each vertex of $\gamma$ is also
a vertex of $\gamma'$. Clearly, with this partial order the set of
graphs becomes a \textit{directed} set.

\begin{figure}
\centerline{\epsfig{file=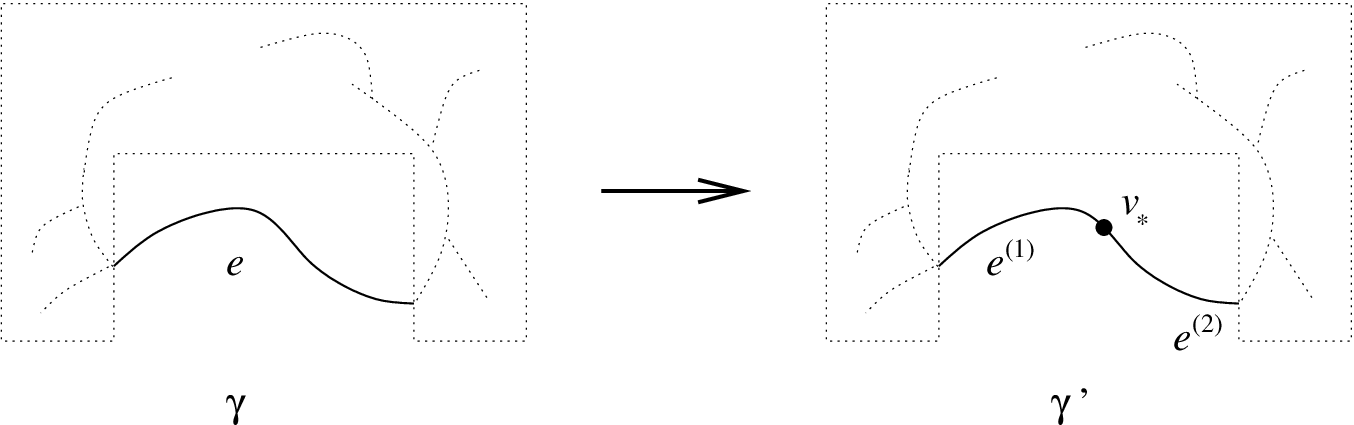, height=4cm}}
\caption{\label{fi1} Operation $\sub_{e,v_*}$ subdividing an edge $e$ 
of a graph}
\end{figure}

\begin{figure}
\centerline{\epsfig{file=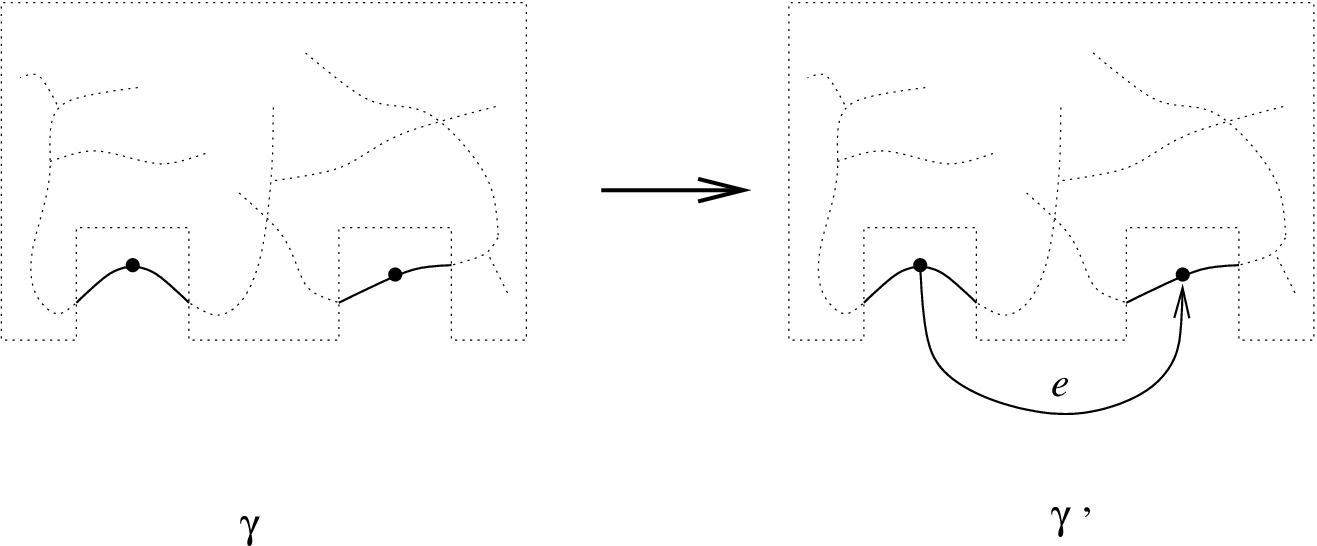, height=5cm}}
\caption{\label{fi2} Operation $\add_e$, adding an edge $e$ to a graph 
  (note that $e$ does not necessarily have to begin and end in
  vertices of $\gamma$)}
\end{figure}

Also note that if
$\gamma'\geq\gamma$, one can obtain $\gamma'$ from $\gamma$ by 
subdividing edges of $\gamma$ and adding further edges. 
Let us denote the graph obtained by subdividing an edge $e$ of a graph
$\gamma$ by adding a vertex $v^*$ by $\sub_{e,v^*}\gamma$ (see figure
\ref{fi1}), and the graph obtained by adding an edge $e$ to $\gamma$
by $\add_e\gamma$ (see figure \ref{fi2}). 

Let consider a smooth principal fiber bundle over $\Sigma$ with
a compact connected structure group $G$, and denote by $\con$ the
space of smooth connections on this bundle.
It turns out to be convenient to consider a slightly more general 
class of functions on $\con$ than the holonomies \eqref{eq1.2}:
\begin{defi}
\label{de2.1}
    A function $c$ depending on connections $A$ on $\Sigma$ just in
    terms of their holonomies along the edges of a graph, i.e. 
    \begin{equation*}
      c[A]\equiv c(h_{e_1}[A],h_{e_2}[A],\ldots,h_{e_n}[A] ), \qquad
      e_1,e_2,\ldots, e_n\quad\text{ edges of some }\gamma, 
    \end{equation*}
    where $c(g_1,\ldots, e_n)$, viewed as a function on $G^n$, is
    \textit{continuous}, will be called \textit{cylindrical}. 
\end{defi}
Now, given a graph $\gamma$, any connection of the principal fiber bundle over $\Sigma$ gives rise to a connection over (the union of) the edges of $\gamma$. We denote the space of these connections, modulo gauge transformations that are trivial over the vertices of $\gamma$, as $\con_\gamma$. 
To each of these spaces $\con_\gamma$ there is a
surjective map $\pi_\gamma:\con\longrightarrow\con_\gamma$. Moreover,
these spaces decomposes into the Cartesian product 
\begin{equation*}
  \con_\gamma = \underset{e\in E(\gamma)}{\times} \con_e.
\end{equation*}
There is a bijection between the elements of $\con_e$ and parallel transport maps between the fibers over the beginning and endpoints of $e$, 
and upon fixing an element of those fibers, there are bijections
$\Lambda_e:\con_e\longrightarrow G$. Obviously these bijections are not
unique. Combining the maps $\Lambda_e$ of all edges of a graph $\gamma$, we also obtain a bijection $\Lambda_\gamma:\con_\gamma\longrightarrow G^{|\gamma|}$.  

Finally, whenever $\gamma'\geq\gamma$ there is a projection map
$p_{\gamma\gamma'}:\con_{\gamma'}\longrightarrow\con_{\gamma}$  
such that $\pi_\gamma=p_{\gamma\gamma'}\circ\pi_{\gamma'}$. 
The spaces $\{\con_\gamma\}$ together with the maps
$\{p_{\gamma\gamma'}\}$ form a \textit{projective
  family}. Consequently, there is a space $\gcon$, the projective limit 
of the projective family, containing all the $\con_\gamma$ with the
appropriate inclusion relations implied by the projections
$p_{\gamma\gamma'}$. The cylindrical functions of Definition \ref{de2.1}
extend to functions on $\gcon$ in a natural way. Their
closure with respect to the sup-norm is an Abelian C* algebra which is usually
denoted by $\overline{\cyl}$. Its spectrum can be identified with $\gcon$, thus
endowing it with a Hausdorff topology in which it is a compact space. 
This shows that $\gcon$ is the natural home for the 
cylindrical functions. Moreover, as a consequence of the 
the Riesz-Markov Theorem, cyclic representations of $\cyl$ are in one
to one correspondence with positive Baire measures on $\gcon$. 

The projective techniques yield an elegant characterization of measures 
on $\gcon$: On the one hand, a measure $\mu$ gives rise to a family of 
measures $\{\mu_\gamma\}$ where $\mu_\gamma$ is a measure on
$\con_\gamma$ by pushing $\mu$ forward with the maps
$\pi_\gamma$. The measures $\{\mu_\gamma\}$ bear \textit{consistency
  relations} among each other: Whenever $\gamma'\geq\gamma$ one has
\begin{equation*}
  p_{\gamma\gamma'*}\mu_{\gamma'}=\mu_{\gamma}. 
\end{equation*}
On the other hand, it was shown in \cite{Baez:1994id} that also the converse holds
true: Every consistent family $\{\mu_\gamma\}$ of measures on the
$\con_\gamma$ gives rise to a measure $\mu$ on $\gcon$. Moreover,
properties of the measures translate between these two presentations:
If the measure $\mu$ is normalized, so are the measures
$\{\mu_\gamma\}$ and vice versa. If the measure $\mu$ is positive 
so are the measures $\{\mu_\gamma\}$ and vice versa.

Finally note that via the maps $\Lambda_\gamma$ described above, the family
$\{\mu_\gamma\}$ can be pushed forward to obtain a family of measures 
on Cartesian products of $G$. We will denote these measures by
$\{\tilde{\mu}_\gamma\}$. Everything said about the relation between 
$\mu$ and $\{\mu_\gamma\}$ certainly also holds for $\mu$ and
$\{\tilde{\mu}_\gamma\}$.  
%Let $F$ be cylindrical on $\gamma$ and 
%\begin{equation}
%\label{eq2.1}
%  \int_{\gcon}F\, d\mu=\int_{G^{\betr{E(\gamma)}}}(\Lambda_\gamma^{-1})_*
%F(g_1,\ldots, g_{\betr{E(\gamma)}})
%\,d\mu_\gamma(g_1,\ldots, g_{\betr{E(\gamma)}})
%\end{equation}

In all of the following we will restrict ourselves to a certain
subclass of measures on $\gcon$: We just consider measures $\mu$ such 
that 
\begin{equation}
\label{eq2.2}
d\widetilde{\mu}_\gamma(g_1,\ldots, g_{\betr{E(\gamma)}})=
f_\gamma(g_1,\ldots, g_{\betr{E(\gamma)}})\,\, d\mh(g_1)\ldots
d\mh(g_{\betr{E(\gamma)}}),  
\end{equation}
where $\mh$ is the Haar measure on $G$. Thus we exclude measures 
whose cylindrical projections pushed forward to $G^{|\gamma|}$, $(\Lambda_\gamma)_*\mu_\gamma$, 
would also contain a pure point part and a part singular with respect
to the product of Haar measures on $G$.   
So in all of the following we will consider measures which can be
characterized in terms of a family $\{f_\gamma\}$ of functions
\begin{equation*}
  f_\gamma: G^{\betr{E(\gamma)}}\longrightarrow \C
\end{equation*}
via \eqref{eq2.2}.
Note that the
representation in terms of these functions 
does in general depend on the choice of identifications
$\{\Lambda_e\}$. The dependence is however, a rather simple one. To simplify the notation, in the following we will assume a fixed choice of 
$\Lambda_\gamma$, and 
do not explicitly distinguish anymore between families $\{\mu_\gamma\}$
and $\{\widetilde{\mu}_\gamma\}$. Note also that there is no dependence on the maps $\Lambda_\gamma$ in the case that the measure $\mu$ is gauge invariant. 

Let $\{f_\gamma\}$ be a family of functions defining a positive,
normalized measure $\mu$ on $\gcon$ by way of \eqref{eq2.2}. 
Then positivity implies    
\begin{equation}
\tag{pos}
\label{pos1}
f_\gamma\geq 0 \qqquad \text{pointwise on }G^{\betr{E(\gamma)}},  
\text{ for all }\gamma.
\end{equation}
Normalization implies
\begin{equation}
\tag{norm}
\label{norm1}
\int_G f_e(g)\,d\mh(g)=1 \qqquad \text{for all edges }e.  
\end{equation}
Consistency implies 
\begin{align}
\tag{add}
\label{add1}
f_{\gamma}(g_{e_1},\ldots,g_{e_n})&=
\int f_{\add_{e}(\gamma)}(g_e,g_{e_1},\ldots,g_{e_n})\, d\mh(g_e),\\
\tag{sub}
\label{sub1}
f_{\gamma}(g_{e_1},\ldots,g_{e_i},\ldots,g_{e_n})&=
\int f_{\sub_{e_i}(\gamma)}(g_{e_1},\ldots,g,g^{-1}g_{e_i},\ldots, g_{e_n})\,
d\mh(g).
\end{align}
It is easy to check, that also the converse holds true:
\begin{prop}
  Let a family $\{f_\gamma\}_\gamma$ of functions $f_\gamma$ on 
$G^{\betr{E(\gamma)}}$ be given that fulfills \eqref{pos1},
\eqref{norm1}, \eqref{add1} and \eqref{sub1}. Then this family defines 
a positive normalized measure $\mu$ on $\gcon$ by virtue of  
\eqref{eq2.2}. 
\end{prop}
Let us close this section by pointing out that a wide variety of
measures on $\gcon$ has been constructed with the projective
techniques reviewed above. Most important is perhaps the
Ashtekar-Lewandowski measure $\mal$ \cite{Ashtekar:1995mh} 
which is obtained by setting all
$f_\gamma$ equal to 1. Other measures are the
diffeomorphism invariant Baez measures \cite{Baez:1993wg}, the heat kernel
measure of \cite{Ashtekar:1995wa}, the $r$-Fock measures 
constructed from the Gaussian measure of the free electromagnetic
field \cite{V2}, and the measures obtained with the complexifier 
method \cite{Thiemann:2002vj}. 
%---------------------------------------------------------------------------
\section{Admissibility}
%---------------------------------------------------------------------------
\label{se3}
Up to now we have only considered representations of the algebra
$\cyl$ of cylindrical functions. Now we will turn to the momentum
observables $E_{S,f}$ defined in \eqref{eq1.1}. 
First, we should note that to avoid pathologies, it is required to put 
certain restrictions on the surfaces $S$ to be considered.  
In the following, we will always assume that the surfaces $S$ are analytically
embedded in $\Sigma$, simply connected and such that $S=S-\partial S$.   
We caution the reader that we will not always explicitly state this in the
following. Also let us restrict the vector
fields $f$ used in the definition of the $E_{S,f}$ to be smooth and bounded. 
Under these assumptions, one can compute the Poisson brackets for the 
$c\in \cyl$ with the $E_{S,f}$ \cite{Ashtekar:1997eg}: 
\begin{equation}
\label{eq3.2}
  \{E_{S,f},c\} = X_{S,f}[c], \quad\text{ where }\quad
X_{S,f}[c]=\frac{\kappa}{2}\sum_{v \in S\cap \gamma}
\,\sum_{e\in E(v)} \sigma(v,e) f_i({v}) X^i_e[c].
\end{equation}
In this formula we have assumed without loss of generality that all
the intersections of $\gamma$ and $S$ are vertices of
$\gamma$. Moreover, $X_e$ denotes the right or left invariant
vector-field on $\sutwo$, depending on whether $e$ is ingoing 
or outgoing, respectively, acting on
the entry corresponding to $e$ of $c$ written as a function 
on $\sutwo^{\betr{E(\gamma)}}$. Finally, $\sigma(v,e)$ is the sign 
of the natural pairing between orientation two-form on $S$ and 
tangent of $e$ in $v$ (and 0 if $e$ is tangential). $\kappa$ is the
coupling constant of gravity.  

In the present section we are going to consider the following problem: 
Given a measure $\mu$ on $\gcon$, what are the conditions $\mu$ has to 
satisfy in order to allow for a representation of the $E_{S,f}$ on the 
Hilbert space $\mathcal{H}=L^2(\gcon, d\mu)$ by selfadjoint operators?

Since the operators representing the $E_{S,f}$ will in general be
unbounded, it is necessary to make the notion ``representation'' in
the question formulated above a bit more precise by putting some 
requirement on the domains of the operators representing the $E_{S,f}$.  
In \cite{1} we have argued that a reasonable requirement is that the
smooth cylindrical functions 
$\cyl^\infty$ be in those domains. Let us adopt this requirement
and cite from \cite{1} a simple criterion for a measure $\mu$ to carry such a
representation:
\begin{prop}
\label{pr4}
  Let a positive measure $\mu$ on $\gcon$ be given. 
  Then a necessary and sufficient condition for the existence 
  of a representation of the $E_{S,f}$ on $\mathcal{H}=L^2(\gcon,
  d\mu)$ by symmetric operators with domains containing $\cyl^\infty$ 
  is that derivations $X_{S,f}$ map differentiable cylindrical functions that are zero almost everywhere  with respect to $\mu$ to functions that are again zero $\mu$-a.e.\ (condition (nul) of \cite{1}), that they extend to $\cyl^\infty$ as well-defined operators, and that for
  each surface $S$ and co-vector field $f$ on $S$ there exists a
  constant $C_{S,f}$ such that
  \begin{equation}
    \label{eq6}
    \betr{\Delta_{S,f}(c,\one)} \leq C_{S,f}\norm{c}_{\hilb}\qquad\text{ for 
      all }c\in \cyl^{\infty}
  \end{equation}
  where the sesquilinear form $\Delta_{S,f}$ is given by
  \begin{equation*}
    \Delta_{S,f}(c,c')\doteq \scpr{\rep{X_{S,f}[c]}\one}{\rep{c'}\one}_{\hilb}-
    \scpr{\rep{c}\one}{\rep{X_{S,f}[c']}\one}_{\hilb}, \qquad c,c'\in\cyl.
  \end{equation*}
\end{prop}
In the following, we will also denote by $\Delta_{S,f}(c)$ the anti-linear form $\Delta_{S,f}(c,\one)$.

Let us sketch how this result comes about. The Poisson brackets \eqref{eq3.2}
suggest to represent the $E_{S,f}$ as $\pi(E_{S,f})=i\hbar X_{S,f}$,
since this obviously promotes these brackets to commutation
relations. The problem is that despite the $i$ in the definition of
$\pi$ suggested above, the $\pi(E_{S,f})$ will in general 
not be symmetric, since the measure can have a non-vanishing
``divergence'' with respect to the vector fields $X_{S,f}$,
i.e. formally
\begin{equation*}
  i\hbar X_{S,f}[d\mu]\neq 0. 
\end{equation*}
Certainly this equation does not make sense as it
stands. However, the form $\Delta_{S,f}$ defined in 
Proposition \ref{pr4} is the appropriate definition for this 
divergence. The condition on $\mu$ exhibited in Proposition \ref{pr4}
is simply the requirement that $\Delta_{S,f}$ be given by an $L^2$
function, $F_{S,f}$, say. If this is the case, we can represent
$E_{S,f}$ as $\pi(E_{S,f})=i\hbar X_{S,f}+\hbar F_{S,f}/2$, which, as
can be easily checked, is symmetric. 

Before we proceed, let us make two remarks: The first one is that a
priory it is only necessary to require the ``divergences''
$\Delta_{S,f}$ to be operators on $\hilb$, i.e. they do not have to be 
square integrable. As soon as one requires $\cyl^\infty$ to be part of 
the domain, they are
automatically $L^2$. The second remark we want to make is that it was
realized already in \cite{Ashtekar:1995wa}, that when considering more general
measures then $\mal$, a divergence term will have to be added to the
$X_{S,f}$ to make them symmetric. The requirement of \textit{compatibility} 
between measure and vector-field used there, seems too restrictive,
however, since it implies that the divergence is a cylindrical
function. 

Let us call a surface $S$ \textit{admissible} with respect to a
positive measure $\mu$, if the action of the derivations $X_{S,f}$ can be extended to equivalence classes under the measure $\mu$, and if furthermore
the $\Delta_{S,f}$ are in $L^2(\gcon,
d\mu)$ for all smooth co-vector fields $f$. Then the following is a
simple corollary  of Proposition \ref{pr4}:
\begin{prop}
\label{pr5}
  A surface $S$ is admissible with respect to a positive measure
  $\mu$, coming from a family $\{f_\gamma\}$ via \eqref{eq2.2} iff
for all co-vector fields $f$ there is a constant $C_f$ such that 
\begin{equation*}
  \norm{X_{S,f}[\ln f_\gamma]}_{L^2(\con_\gamma, d\mu_\gamma)}\leq C_{S,f}\qquad \text{ for all graphs }\gamma, 
\end{equation*}
 where the $f_\gamma$ are the functions characterizing $\mu$ according 
 to \eqref{eq2.2}.
\end{prop}
\begin{proof}
  Assume $S$ to be admissible with respect to $\mu$, and let $F$ be
  cylindrical on $\gamma$. Then there are constants $C_{S,f}$ such
  that 
  \begin{equation*}
    \betr{\Delta_{S,f}(F)}=\betr{\int \overline{F} X_{S,f}[\ln
      f_\gamma]f_\gamma\,d\mh^{\betr{E(\gamma)}}}\leq C_{S,f} \norm{F}_\mu 
    =C_{S,f} \norm{F}_{L^2(\con_\gamma, d\mu_\gamma)},
  \end{equation*}
 so $X_{S,f}[\ln f_\gamma]$ is in $L^2(\con_\gamma,d\mu_\gamma)$. This 
 allows us to plug it into $\Delta_{S,f}$, yielding
 \begin{equation*}
   \norm{X_{S,f}[\ln f_\gamma]}_\mu^2=\betr{\Delta_{S,f}(X_{S,f}[\ln
     f_\gamma])}\leq C_{S,f}\norm{X_{S,f}[\ln f_\gamma]}_\mu
 \end{equation*}
whence $\norm{X_{S,f}[\ln f_\gamma]}_{\mu_\gamma}\leq C_{S,f}$, independently
of $\gamma$. 

Vice versa, assume that there are constants $C_{S,f}$, such that
$\norm{X_{S,f}[\ln f_\gamma]}_\mu\leq C_{S,f}$, independently 
of $\gamma$. Then for $F$ cylindrical on $\gamma$, 
\begin{align*}
\betr{\Delta_{S,f}(F)}&=\betr{\int \overline{F} X_{S,f}[\ln
      f_\gamma]f_\gamma\,d\mh^{\betr{E(\gamma)}}}
      =\betr{\scpr{F}{X_{S,f}[\ln f_\gamma]}_{L^2(\con_\gamma,d\mu_\gamma)}}\\
&\leq \norm{F}_{L^2(\con_\gamma,d\mu_\gamma)}
\norm{X_{S,f}[\ln f_\gamma]}_{L^2(\con_\gamma,d\mu_\gamma)}\\
&\leq C_{S,f}\norm{F}_{L^2(\gcon,d\mu)}, 
\end{align*}
independently of $\gamma$. 
\end{proof}
%-----------------------------------------------------------------------------
\section{Admissibility in the $\uone$ case}
%-----------------------------------------------------------------------------
\label{se9}
Up to now, our considerations applied either to a general compact
connected gauge group, or at least to $\sutwo$. From now on, we will 
turn our attention to the analogous, but technically much simpler case 
$G=\uone$, i.e. the electromagnetic field. 
We caution the reader, that from now on, all measures are assumed to 
be obtained from families of functions $\{f_\gamma\}$ via \eqref{eq2.2},
without explicitly stating it. 

Let us start 
by introducing some notation. We will be 
very brief and refer to \cite{V2} for more thorough information on the  
$\uone$ theory. 
Let us parametrize $\uone$ as $g(\varphi)=\exp{i\varphi}$ where $\varphi$ is in 
$[0,2\pi]$. The Haar measure is then simply given by $d\varphi/2\pi$.
The irreducible representations are labeled by $n\in\Z$ and are
parametrized by $\pi_n(g(\varphi))=\exp{in\varphi}$.
 
In the following, we will denote by $\gcon_{\uone}$ the space of
generalized $\uone$ connections. 
The \textit{charge network functions} $T_{\gamma,\avec{n}}$ on 
$\gcon_{\uone}$ are defined as   
\begin{equation*}
  T_{\gamma,\avec{n}}
  (\varphi_1,\ldots,\varphi_{\betr{E(\gamma)}})= e^{in_1\varphi_1}\ldots 
  e^{in_{\betr{E(\gamma)}}\varphi_{\betr{E(\gamma)}}}. 
\end{equation*} 
They form an orthonormal basis in $L^2(\gcon_{\uone}, d\mal)$. 
Let us also introduce their integrals with respect to a measure $\mu$ 
\begin{equation*}
  f^{(\avec{n})}_\gamma:=\int \overline{T_{\gamma,\avec{n}}}\, d\mu.  
\end{equation*}
Since $f^{(\avec{n})}_\gamma$ is nothing else then a specific Fourier
coefficient of the function $f_\gamma$, we will call the family 
$\{f^{(\avec{n})}_\gamma\}$ the \textit{Fourier coefficients of the
  measure $\mu$}. The requirements \eqref{norm1}, \eqref{pos1},
\eqref{sub1}, \eqref{add1} have straightforward analogs in the family 
$\{f^{(\avec{n})}_\gamma\}$. 
%We note however that not all families
%$\{f^{(\avec{n})}_\gamma\}$ fulfilling those requirements do indeed define
%measures on $\gcon_{\uone}$, due to the fact that there are $L^1$
%functions on $[0,2\pi]$ whose Fourier series does not converge. 
We furthermore note that $\norm{T_{\gamma,\avec{n}}}_\mu=1$ for any
normalized $\mu$.   

For the $\uone$ theory, the co-vector-fields $f$ in the definition of
$E_{S,f}$ are just functions, so we can simplify notation by replacing 
them all by 1. 
The action of the vector-fields $X$ on the charge network functions then read
\begin{equation*}
X_{S}[T_{\gamma,\avec{n}}] 
=\frac{\kappa}{2}\bigg(\sum_{e\in E(\gamma)} 
I(S,e)n_e \bigg) T_{\gamma,\avec{n}},  
\end{equation*}
where $I(S,e)$ is the \textit{signed intersection number} of $e$ and $S$
which we define as follows: Call an intersection of an edge $e$ and a
surface $S$ \textit{proper} if it is not the start or endpoint of $e$
and $e$ is transversal to $S$ at the intersection. Let $P_\pm$ be the
number of proper intersections of $S$ and $e$ with positive/negative
relative orientation of $S$ and $e$ at the intersection point and 
$I_\pm$ the number of intersections with positive/negative
relative orientation that are not proper. Then 
$I(S,e)=P_+-P_-+(I_+-I_-)/2$. 
Finally, we note the following useful formula:
\begin{equation}
\label{eq3.3}
  \Delta_S(T_{\gamma,\avec{n}})= \frac{\kappa}{2}\bigg(\sum_{e\in E(\gamma)} 
I(S,e)n_e \bigg) f_\gamma^{(\avec{n})}. 
\end{equation}

We will now examine more closely the admissibility of surfaces with
respect to measures on $\gcon_{\uone}$. We will see that 
the reasons for surfaces not to be admissible can be manifold:
Firstly, note that since the representation labels $n_e$ in 
\eqref{eq3.3} can be arbitrary large, admissibility requires that the
higher Fourier components of the measure have to be suitably damped. 
This is the reason why the $r$-Fock measures do not have
admissible surfaces, as we will show below.
 
Another reason for non-admissibility is that the vector fields
$X_S$ act on cylindrical functions as sums of
derivatives. The number of terms in this sum can be very large when
the intersections of graph and surface become numerous. Therefore 
to allow for admissible surfaces, $f_\gamma$ must contain sufficient
information about the geometry of the edges contained in $\gamma$ to
tell how many times an edge can intersect with a given surface. 
This is not possible if the measure is required to be
diffeomorphism invariant -- we will prove below that, with the
exception of $\mal$, such measures do not have any admissible 
surfaces.
 
Finally, for the same reason $f_\gamma$ has to contain information
about the positions of the edges of $\gamma$ relative to each
other. We will see below that this forces \textit{factorizing
  measures}, i.e. measures for which the $f_\gamma$ factorize into a
product of functions just depending on single edges, to be extremely
close to $\mal$ in a certain sense, if they are to allow for
admissible surfaces. 
\\
\\
\\
%---------------------------------------------------------------------------
{\bf Diffeomorphism invariant measures.}
%---------------------------------------------------------------------------
\\
\\
Of special importance for quantum gravity are measures that do not
depend on any geometric background structures (such as a metric or a
connection) on $\Sigma$. The requirement of background independence
can be formalized as follows: Analytic diffeomorphisms $\phi$ naturally act
on the space of graphs by mapping a graph $\gamma$ to its image
$\phi(\gamma)$ which clearly is a graph again. Consequently, they also 
act on cylindrical functions by
\begin{equation*}
 F[A]\equiv F(h_{e_1}[A],\ldots,h_{e_n}[A])\longmapsto
 F(h_{\phi(e_1)}[A],\ldots,h_{\phi(e_n)}[A])=:\phi(F)[A].
\end{equation*}
A measure $\mu$ is called invariant under analytic diffeomorphisms, or 
shorter, \textit{diffeomorphism invariant}, if 
\begin{equation*}
  \int F\, d\mu = \int \phi(F)\, d\mu
\end{equation*}
for all $F\in\cyl$ and all analytic diffeomorphisms $\phi$. A measure
coming from a family $\{f_\gamma\}$ of functions in the sense of
\eqref{eq2.2} is clearly  
diffeomorphism invariant iff $f_\gamma=f_{\phi(\gamma)}$ for all
graphs and all analytic diffeomorphisms $\phi$. 

Examples of diffeomorphism invariant measures are the Baez measures
\cite{Baez:1994id} and $\mal$. 
A bit surprisingly, it turns out that at least in
the $\uone$ case considered here, $\mal$ is the only such measure that 
has admissible surfaces. This shows again that $\mal$ is a very
special measure. 
\begin{prop}
\label{pr1}
Let $\mu$ be a diffeomorphism invariant normalized measure on
$\gcon_{\uone}$ coming from a family $\{f_\gamma\}$ of functions in
the sense of \eqref{eq2.2}. If there exists an open, simply connected
surface admissible with respect to $\mu$, then $\mu=\mal$.    
\end{prop}
Before proving this Proposition, we have to provide a rather technical 
Lemma: 
\begin{lemm}
\label{le1}
Given a graph $\gamma$, an open, simply connected surface $S$ and a vector 
$\avec{m}=(m_1,\ldots,
m_{\betr{E(\gamma)}})\in \Z^{\betr{E(\gamma)}}$, 
there is an analytic
diffeomorphism $\phi^A_{\avec{m}}$ of $\Sigma$ such that
\begin{equation*}
I(S,\phi^A_{\avec{m}}(e_1))=m_1,\quad\ldots\quad 
I(S,\phi^A_{\avec{m}}(e_{\betr{E(\gamma)}}))=m_{\betr{E(\gamma)}},
\end{equation*}
where $e_1,\ldots,e_{\betr{E(\gamma)}}$ are the edges of $\gamma$. 
\end{lemm}
\begin{proof}
Let us start by considering a single edge $e$. The first observation 
we make is that for arbitrary $m\in\Z$ there is a smooth
diffeomorphism $\phi_m$ of $\Sigma$ such that
$I(S,\phi_m(e))=m$. For example, this diffeomorphism might ``drag
out'' some part of $e$ to create the desired intersections but be
the identity far away from $e$ (see figure \ref{fi3}).    
\begin{figure}
\centerline{\epsfig{file=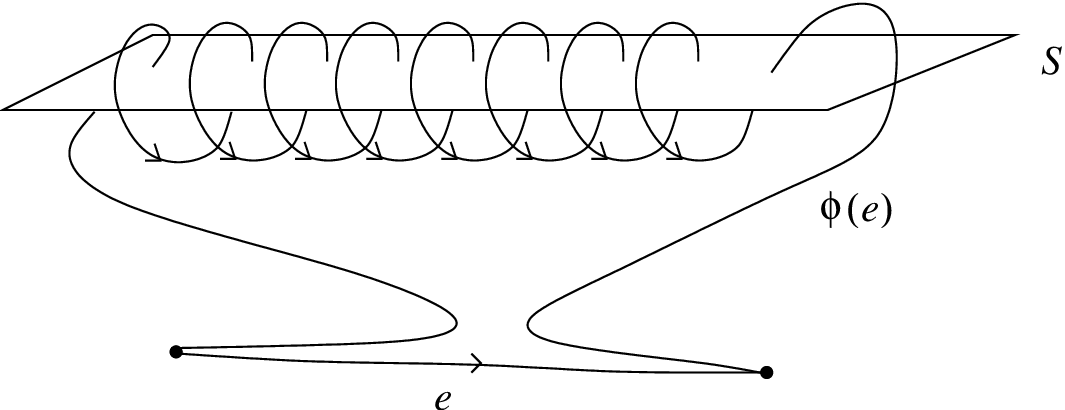, height=4cm}}
\caption{\label{fi3} An example for $e$ and $\phi(e)$}
\end{figure}
Note that such a diffeomorphism exists, whether $e$ is a loop or not,
that its existence depends however on just admitting surfaces that are
open and simply connected.  

The problem is that $\phi_m$ will in general not be
analytic. Therefore we have to establish now that there is even an analytic 
diffeomorphism $\phi_m^A$ which does the same job, i.e. for which 
$I(S,\phi^A_m(e))=m$. We will do this by using powerful mathematical
results on real analytic manifolds and analytic approximations to
diffeomorphisms. 

Note first that since $S$ does not contain its boundary, 
all intersections of $\phi_m(e)$ with $S$ happen in the inside of
$S$. Therefore there is a tube $T$ containing $\phi_m(e)$ such that
every other curve starting at the starting point and ending at the
endpoint of $\phi_m(e)$ and lying entirely inside $T$ will also have
the signed intersection number $m$ with $S$ (see figure \ref{fi4}). 
\begin{figure}
\centerline{\epsfig{file=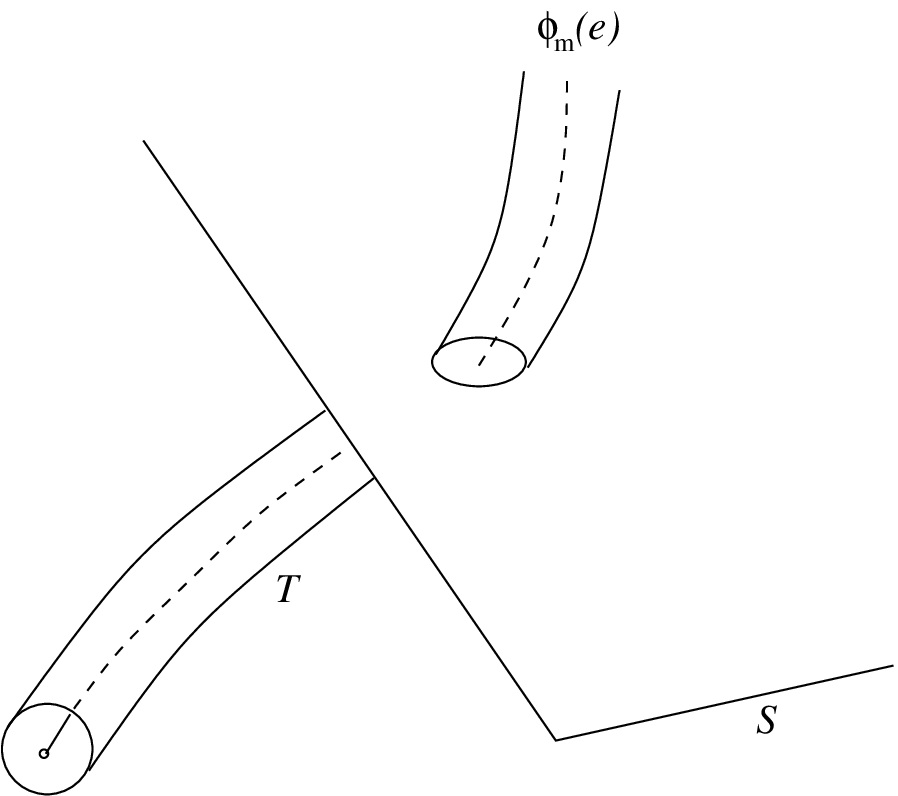, height=5cm}}
\caption{\label{fi4} The tube $T$}
\end{figure}
Therefore, if we manage to map $e$ into $T$ such that the endpoints of 
the image are close to that of $\phi_m(e)$, we are done. 

Let us first recall the famous result by Grauert \cite{grauert} that
every real analytic paracompact manifold can be analytically embedded
into $\R^n$ for some large enough $n$. Therefore we will, without loss
of generality, regard $\Sigma$ as embedded into $\R^n$ in this way. 

Now $R^n$ comes with a globally defined metric $\rho(p,p')=\norm{p-p'}$ and 
induces a metric tensor on $\Sigma$ that locally on $\Sigma$ 
gives rise to a metric $\rho'$. As $\Sigma$ is embedded, the topology
induced by $\rho$ on $\Sigma$ and that given by $\rho'$ coincide. That said,
we turn to a powerful result by Huebsch and Morse \cite{huebsch} 
on approximations of diffeomorphisms of manifolds analytically 
embedded in $\R^n$: Theorem 1.4 of \cite{huebsch}, slightly
specialized to our needs, reads
\begin{theo}[Theorem 1.4 of \cite{huebsch}] 
Let $\Sigma$ be analytically embedded in $\R^n$, $\phi$ a smooth
diffeomorphism of $\Sigma$ and $\delta>0$. Then there is an analytic
diffeomorphism $\phi^A$ of $\Sigma$ that satisfies
\begin{equation*}
  \rho(\phi(p),\phi^A(p))\leq \delta.
\end{equation*}
\end{theo}
That is, we can approximate $\phi_m$ by an analytic diffeomorphism,
the quality of the approximation being measured by $\rho$. What remains to be
shown is that this also gives an approximation controlled by $\rho'$.
To this end, let us parameterize $e$ by $t\in [0,1]$ and let 
$\epsilon$ be chosen such that the $\epsilon$-balls 
$U_\epsilon(e(t))$ wrt. $\rho'$ in $\Sigma$ lie within the tube $T$
for all parameter-values $t$. Since the topology coming from $\rho'$
and that induced by $\rho$ on $\Sigma$ coincide, we can find $\delta(t)>0$
such that the intersection of the $\delta$-ball $U_{\delta(t)}(e(t))$
in $\R^n$ with $\Sigma$ lies within $U_\epsilon(e(t))$. Since the
embedding of $\Sigma$ and the metric tensors are all continuous, we
can choose $\delta(t)$ continuous. Letting $\delta$ be the (nonzero) 
minimum of $\delta(t)$, we can invoke the above theorem to find
$\phi^A_m$ which indeed sends all the points of $e$ into $T$, 
the endpoints of 
the image being  close to those of $\phi_m(e)$. Therefore
$I(S,\phi^A_m(e))=m$, and we have proven the theorem for the case of a
single edge. 

A similar reasoning shows the existence of an analytic diffeomorphism 
$\phi^A_{\avec{m}}$, mapping a given graph $\gamma$ to one such that
$I(S,\phi^A_{\avec{m}}(e_1))=m_1$, $I(S,\phi^A_{\avec{m}}(e_2))=m_2,\ldots$:
Again it is not hard to see that there is a smooth diffeomorphism doing
the job and that it can be suitably approximated by an analytic
diffeomorphism. Since no new idea but just a lot more notation is needed
in this case, we refrain from giving the details. 
\end{proof}
With the above Lemma at hand, the proof of the Proposition is now
straightforward:
\begin{proof}[Proof of Proposition \ref{pr1}]
Let $\mu$ be a diffeomorphism invariant normalized measure on 
$\gcon$ and $S$ a surface which is admissible with respect to $\mu$. 
Pick an arbitrary graph $\gamma$ and a vector $\avec{m}\in
\Z^{\betr{E(\gamma)}}$ and denote the analytic diffeomorphism 
provided by Lemma \ref{le1} by $\phi_{\avec{m}}$.
One computes
\begin{equation*}
  \betr{\Delta_S(T_{\avec{n},\phi_{\avec{m}}(\gamma)})}=\hbar\frac{\kappa}{2}
\betr{\sum_I n_I m_I}\betr{f_{\phi_{\avec{m}}(\gamma)}^{(\avec{n})}}=
\hbar\frac{\kappa}{2} \betr{\sum_I n_I m_I}\betr{f_\gamma^{(\avec{n})}}  
\end{equation*}
where the last equality is due to the diffeomorphism invariance of the 
measure. Since $S$ is assumed to be admissible, this has to be bounded 
independently of $\avec{n}$. But as $\avec{m}$ is arbitrary,
$\betr{\sum_I n_I m_I}$ can be made arbitrarily large whenever  
$\avec{n}\neq 0$. So in this case $f_\gamma^{(\avec{n})}$ has to be
zero. $f_\gamma^{(0)}$ has to be 1 due to normalization. Since 
$\gamma$ was arbitrary, we have shown that indeed $\mu=\mal$. 
\end{proof}
%---------------------------------------------------------------------------
{\bf Factorizing measures.}
%---------------------------------------------------------------------------
\\
\\ 
Let $\{f_\gamma\}_\gamma$ be a family defining a positive normalized
measure on $\gcon_{\uone}$. We will call this measure
\textit{factorizing} if
\begin{equation*}
f_\gamma(g_1,\ldots,g_{\betr{E(\gamma)}})=\prod_{i=1}^{\betr{E(\gamma)}}
f_{e_i}(g_i) \qquad{ where }\qquad E(\gamma)=\{e_1,\ldots,e_{\betr{E(\gamma)}}\}.   
\end{equation*}
Examples for factorizing measures are the heat kernel measures  
and $\mal$. Factorizing measures are particularly easy to deal with,
and regarding admissibility, we find the following 
\begin{prop}
\label{pr2}
  If $\mu$ is a positive, normalized, factorizing measure on $\gcon_{\uone}$ 
  (defined by a family of functions 
  $\{f_e\}$) and $S$ is an admissible surface with respect to
  $\mu$. Then, of all the edges $e$ intersecting $S$ once and with a
  given relative orientation, at most countably infinitely many can 
  have $f_e\neq 1$.  
\end{prop}
This result is a bit technical, but it has interesting consequences,
for example for Euclidean
  invariant measures: If $\Sigma=\R^3$, we call a measure $\mu$ 
\textit{Euclidean invariant} if for all cylindrical functions $F$ and
all Euclidean transformations $T$
\begin{equation*}
  \int F\, d\mu = \int T(F)\, d\mu.
\end{equation*}
A consequence of Proposition \ref{pr2} is
\begin{coro}
\label{co1}
 Let $\mu$ be an Euclidean invariant positive, normalized, and
 factorizing measure on
 $\gcon_{\uone}$, possessing an admissible surface. Then $\mu=\mal$.   
\end{coro}
Let us first prove the Proposition.
\begin{proof}[Proof of Proposition \ref{pr2}] 
Consider a surface $S$ and a graph $\gamma$ with $N$ edges 
such that all edges are intersecting $S$ exactly once, and with the
same relative orientation (see figure \ref{fi9}).
\begin{figure}
\centerline{\epsfig{file=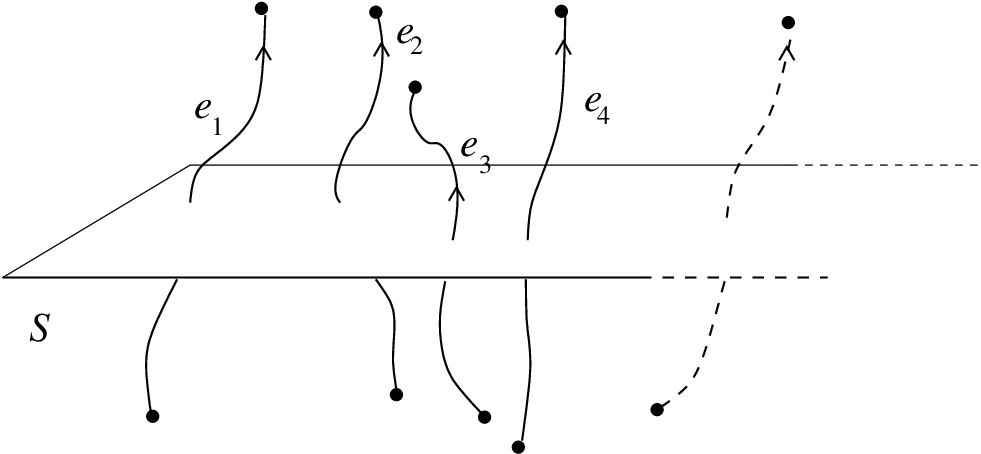, height=3cm}}
\caption{\label{fi9} Surface and graph considered in the proof of
  Proposition \ref{pr2}}
\end{figure}
An easy computation shows that
\begin{equation*}
f_\gamma^{-1}(\avec{\varphi})\betr{X_{S}[f_\gamma]}^2=
\sum_{I=1}^N 
\betr{\frac{f'_{e_I}(\varphi_{e_I})}{f_{e_I}(\varphi_{e_I})}}^2\prod_{K=1}^N
f_{e_K}(\varphi_{e_K}) 
+\sum_{I\neq J}\frac{\overline{f'_{e_I}}(\varphi_{e_I})}{\overline{f_{e_I}}(\varphi_{e_I})}
\frac{{f'_{e_J}}(\varphi_{e_J})}{{f_{e_J}}(\varphi_{e_J})}\prod_{K=1}^N f_{e_K}(\varphi_{e_K}). 
\end{equation*}
Using \eqref{norm1} and the symmetry properties of the derivatives with
respect to the Haar measure, we see that the integral over the second
term vanishes, and therefore
\begin{equation}
\label{eq3.1}
\int_{\uone^N} f_\gamma^{-1}\betr{X_{S}[f_\gamma]}^2\, d\mh^N
=\sum_{I=1}^N \int_0^{2\pi}\frac{1}{f_{e_I}}\betr{\partial_\varphi
  f_{e_I}(\varphi)}^2\,d\varphi.  
\end{equation}
Because of \eqref{pos1}, $1/f_{e_I}$ is strictly positive, so
\begin{equation*}
  \int_0^{2\pi}\frac{1}{f_{e_I}}\betr{\partial_\varphi
  f_{e_I}(\varphi)}^2\,d\varphi \geq 0
\end{equation*}
with equality iff $f_{e_I}(\varphi)= const$. Normedness fixes the constant to be 
1. On the other hand,
because $S$ is assumed to be admissible, Proposition \ref{pr5} requires 
the right hand side of \eqref{eq3.1} to be bounded independently of
$\gamma$. If there would be a more than countably infinite number of
edges $e$ intersecting $S$ once and with the chosen relative
orientation, for which $f_{e}$ is non-constant, there would be
subsequences $e_1, e_2, \ldots$ among these edges such that with 
$\gamma_N:=\cup_{I=1}^N e_N$, $\int
f_\gamma^{-1}\betr{X_{S}[f_\gamma]}^2\, d\mh^N$ would get arbitrarily
large for large $N$. So at most a countably infinite number of these
edges can have $f_e$ non constant.  
\end{proof}
Let us now prove the corollary.
\begin{proof}[Proof of Corollary \ref{co1}]
Let $\mu$ be an Euclidean invariant, positive, normalized, factorizing
measure and $S$ a surface admissible with respect to $\mu$. With
$\{f_e\}$, we denote the family of functions on $\uone$ defining it. 
Consider an arbitrary edge $e$. By Euclidean moves, it can always be
mapped to an edge $e'$ which intersects $S$ at least once. Subdivide   
$e'$ into edges $e'_1,e'_2,\ldots$ such that each of them intersects
$S$ \textit{precisely} once. 
Consider one of those, $e'_I$. By moving it around by Euclidean moves, 
one can obtain an uncountable family of edges intersecting $S$
once. Apply Proposition \ref{pr2} and use Euclidean invariance to
conclude $f_{e'_I}=1$. Do this for all the $e'_1,e'_2,\ldots$. Then
\eqref{sub1} shows that $f_{e'}=1$. Use Euclidean invariance to
finally conclude $f_e=1$.  
\end{proof}
%---------------------------------------------------------------------------
{\bf Varadarajan Measures.}
%---------------------------------------------------------------------------
\\
\\
In \cite{V2} Varadarajan made the remarkable observation that one can 
``import'' measures defined on the space of tempered distributions on
$\R^3$ to $\gcon_{\uone}$. In this subsection, we will consider the
properties of such imported measures with respect to admissibility of
surfaces.  
Let for this purpose be $\Sigma=\R^3$, equipped with the Euclidean
metric. Furthermore denote with $\mathcal{S}$ the Schwarz test
function space on $\R^3$.

We start with the observation that 
\begin{equation*}
h_e[A]=\exp i \int_e A\, ds=\exp i \int_{\R^3}F^{(0)}_e(x)A(x)\,d^3x, 
\end{equation*}
where 
\begin{equation*}
  F^{(0)i}_e(x)=\int_0^1\dot{e}^i(t)\delta(x-e(t))\, dt
\end{equation*}
is the ``distributional formfactor'' of the edge $e$. Using this
notation, the Fourier transforms of a measure $\mu$ can be written as
\begin{equation}
\label{v.1}
  f_\gamma^{(\avec{n})}=\int 
 \exp -i\int_{\R^3}A(x)\left(\sum_I n_I F^{(0)}_{e_I}\right)(x)\,
 d^3x\, d\mu[A].
\end{equation}
An important observation of Varadarajan was that the right hand side 
of \eqref{v.1} formally has the same structure as the 
Fourier transform (or generating functional) 
\begin{equation}
\label{v.5}
  \mathcal{F}(F)=\int \exp -i\bigg(\int_{\R^3}A(x)F(x) d^3x\bigg)\, DA,\qqquad
  F\in \mathcal{S}^3
\end{equation}
for a positive regular Borel measure $DA$ on the space
$(\mathcal{S}^3)'$ of tempered  
distributions. Such measures were extensively studied in quantum field
theory. 
The analogy between \eqref{v.1} and \eqref{v.5} is a priory only
formal, because the
$F^{(0)}_{e_I}$ of \eqref{v.1} are certainly not in
$\mathcal{S}^3$. But Varadarajan realized that on can \textit{define}
measures on $\gcon$ by setting 
\begin{equation}
\label{v.2}
  f_\gamma^{(\avec{n})}=:\int 
 \exp -i\int_{\R^3}A(x)\left(\sum_I n_I F_{e_I}\right)(x)\,
 d^3x\, DA, 
\end{equation}
where now 
\begin{equation}
\label{v.3}
 F_e^i(x):=\int_0^1\dot{e}^i(t)F(x-e(t))\, dt
\end{equation}
for a fixed positive $F\in\mathcal{S}^3$. Consistency of the
$\{f_\gamma^{(\avec{n})}\}$ follows from the behavior of \eqref{v.3}
under composition, and positivity and normedness of the resulting
measure on $\gcon$ from the corresponding properties of the measure
$DA$ on $(\mathcal{S}^3)'$. 

A natural question to ask is whether the resulting measures have
admissible surfaces. The answer is in the negative, as the following
proposition shows:
\begin{prop}
 Any measure obtained from a regular normalized Borel measure on  
 $(\mathcal{S}^3)'$ by Varadarajan's method  as described
 above has no admissible surfaces.   
\end{prop}
\begin{proof}
Let $\mu$ be a measure obtained from a regular Borel measure $DA$ on  
 $(\mathcal{S}^3)'$ as described above, let $S$ be any surface and $e$ 
 an edge intersecting $S$ precisely once. 
 For convenience, let us choose a parametrization of $e$
 such that $e\cap S=e(1/2)$.
 
As a preparation we note that 
\begin{equation*}
  \betr{\Delta_S(T_{e,n})}=\hbar \frac{\kappa}{2} \betr{n}\betr{f^{(n)}_e}=
\hbar \kappa \betr{n}\betr{\mathcal{F}(nF_e)}
\end{equation*}
where $\mathcal{F}$ is the Fourier transform of $DA$ and $F_e$ the form factor
of $e$ as defined above. Now we define the edge $e_\epsilon$ for 
$0<\epsilon< 1$:
\begin{equation*}
 e_\epsilon(t):=e((1-\epsilon)/2+\epsilon t),\qqquad t\in [0,1]. 
\end{equation*}
Thus $e_\epsilon$ becomes shorter and shorter with vanishing
$\epsilon$, but still intersects $S$. Moreover, it is easy to check that 
$F_{e_\epsilon}\longrightarrow 0$ for $\epsilon\longrightarrow 0$, in
the topology of $\mathcal{S}^3$.
 
Now we appeal to the Bochner-Minlos Theorem (see for example
\cite{Glimm:1987ng}) 
which states that the Fourier transform $\mathcal{F}$ of $DA$ is continuous (in
the topology of $\mathcal{S}^3$) and that $\mathcal{F}(0)=1$. Therefore, by
making $\epsilon$ small, we can bring $\mathcal{F}(nF_{e_\epsilon})$ as close to 
1 as we wish. As on the other hand $\betr{n}$ can be arbitrarily
large, we see that $\betr{\Delta_S(T_{e_\epsilon,n})}$ can be made
arbitrarily large. Hence it can not be bounded by a constant
independent of $e_\epsilon,n$ and therefore $S$ can not be
admissible. Since $S$ was completely arbitrary, this completes the
proof.    
\end{proof}
%---------------------------------------------------------------------------
\section{Discussion}
%---------------------------------------------------------------------------
\label{se4}
In the present paper, we have investigated under which circumstances, 
certain ``flux-like'' variables can (not) be represented on measure spaces
over the space of generalized connections. 
This investigation was motivated by recent work on the 
semiclassical sector of loop
quantum gravity. However, many of the results we have obtained concern only 
the simpler case of a $\uone$ gauge theory. 
Thus, an immediate task would be to generalize the results obtained in
this work to other gauge groups, most notably to $\sutwo$. 
However, in view of the results obtained above, 
we have to acknowledge already now that the task of finding
interesting measures supporting the flux operators is more difficult
then on might at first think. 
There are two ways to interpret these difficulties: 
One is to say that background dependent measures lead to a different 
phase of the theory whose ultraviolet behavior is simply not
suited for this kind of observables, so they cease to exist.
A situation vaguely similar is encountered in quantum field theory
on curved spacetimes. There, in some representations of the field
algebra, the operator quantizing the stress energy tensor of the field 
is well defined. In other representations whose
energy content is less well behaved, this operator is not well defined
anymore (elementary discussion and further references in \cite{Wald:1995hf})
The other way to interpret the difficulties is to maintain that 
we have simply not gathered enough
experience with background dependent representations yet, to see how
such measures with admissible surfaces have to be constructed. 

At this point, we are quite dissatisfied with the ratio of
mathematical to physical considerations we produced in the present work. 
Our guess from the experiences with the admissibility condition is
that if measures other than $\mal$ exist which have a sufficiently
large number of admissible surfaces, then we are not likely to find
them by chance or by changing the measures that have already been 
constructed just a little bit.  
Rather, one would need some idea from physics on how to construct such
measures. If on the other hand, such measures do not exist, 
we would like to better understand why this is so from the point of view of a
physicist. The works \cite{V2,V3,Glimm:1987ng} among other things,
take first steps in this direction.  
In any case, many interesting questions are still to be investigated,
and we hope to come back to some of them in future publications. 
\\
\\
\\ 
{\bfseries \Large Acknowledgments}
\\
\\
It is a pleasure to thank Thomas Thiemann for numerous valuable
discussions and suggestions concerning the present work. 
We also thank Jurek Lewandowski for helpful comments on the draft and 
Christian Fleischhack for pointing out a mistake in the proof of Lemma 
\ref{le1} in an earlier version and helping to locate the references
\cite{huebsch} and \cite{grauert}.   
We are also
grateful to Fotini Markopoulou for many very helpful discussions 
on conceptual issues in quantum gravity. 
Financial support from the Studienstiftung des Deutschen Volkes and
the Max Planck-Institut f\"ur Gravitationsphysik are gratefully
acknowledged. Part of this work was also supported by the Spanish MICINN
project No.\ FIS2008-06078-C03-03.
%\bibliographystyle{JHEP-2}
%\bibliography{lqg2}        
\providecommand{\href}[2]{#2}
\begingroup\raggedright

\endgroup

\end{document}